\documentclass[aps,preprint,floats]{revtex4}

\usepackage{graphicx}
\begin{document}
\title{Spiral wave drift in an electric field and scroll wave instabilities}
\author{Herv\'e Henry}
\date{\today}
\affiliation{Center for Theoretical Biological Physics\\ University
of California San Diego\\ 9500 Gilman Dr \\La Jolla CA, USA}
\begin{abstract}
 Here, I first present the numerical computation of  speed and direction  of the drift of  a spiral wave in an excitable medium in the presence of an electric field. In contrast to earlier results, the drift speed presents a strong variation close to the parameter value where the drift speed component along the field direction from parallel becomes anti-parallel. Using a simple phenomenological model and results from  a numerical linear stability analysis of scroll waves, I show this behavior can be attributed to a resonance of the meander modes with the translation modes of the spiral wave. Extending this phenomenological model to scroll waves also clarifies the link between the drift and long wavelength instabilities of scroll waves. 
\end{abstract}
\maketitle

 Spiral waves can be observed in a variety of excitable systems such as Belousov-Zhabotinsky gels\cite{Win72}, colonies of the  \textit{dictyostelium} amoebae\cite{Ara96} and slices  of cardiac tissue\cite{WTBD}. In the latter example, spiral waves of electrical activity have been shown to be the source of  ventricular tachycardia and some of their instabilities are believed to be involved in the transition from tachycardia to  fibrillation, a deadly arrhythmia (For a review see \cite{Fen2002}). This, with the intrinsic interest of those structures,  has lead to an important research effort in order to understand the dynamic and instabilities of spiral waves and of their three dimensional analogous, 
scroll waves.

 In the presence of an external electric field, in the Belousov-Zhabotinsky reaction the center of rotation of spiral waves drift with a speed that presents components both parallel and perpendicular to the applied field \cite{Ste92}. The parallel component of the drift speed was always found to be 
in the direction of the applied field. A numerical study \cite{Kri96} showed 
that depending on the parameter regime the drift direction of the spiral could be either parallel or anti-parallel to the field. The drift of a spiral wave has been linked \cite{hh2000,Hak99} to the curvature instability of scroll waves \cite{Bik94,Alo2003} that leads scroll waves to bend and can finally result in a fibrillation like disordered activity of the medium. It was also  linked to the three dimensional meander of scroll waves \cite{Ara98} which is the three dimensional analog of the meander instability \cite{Kar90,Bar90} characterized by a periodic modulation of the radius of rotation of the spiral wave. 
 This phenomenon has been  studied from a theoretical point of view  \cite{Hak99,Kri96,Zha2003,Wel99}.  However,  most analytical  studies are restricted to the large core  \cite{Hak99} or the small core limit \cite{Mit95} and can not examine the observed change in drift direction. 

In this paper, I present results of numerical computations that show an 
unexpected behavior of the drift speed at the drift direction change. The 
drift speed varies strongly in the vicinity of the transition. 
Using  a reduced model of spiral 
wave in the presence of a small electric field, 
I show that this phenomenon can be attributed to a resonance between meander and translation modes of spiral waves. Finally, using the analogy between the effect of an electric field and the effects of a slight curvature of a scroll wave, I extend this model to scroll waves. This extension brings some light on the link between the drift of a spiral wave in an electric field and long wavelength instabilities of scroll waves and results obtained from this model are in good agreement with the results of the numerical linear stability analysis of scroll waves\cite{hh2000}.

\begin{figure}
\begin{center}
\begin{tabular}{cccc}
\includegraphics[width=0.5\textwidth]{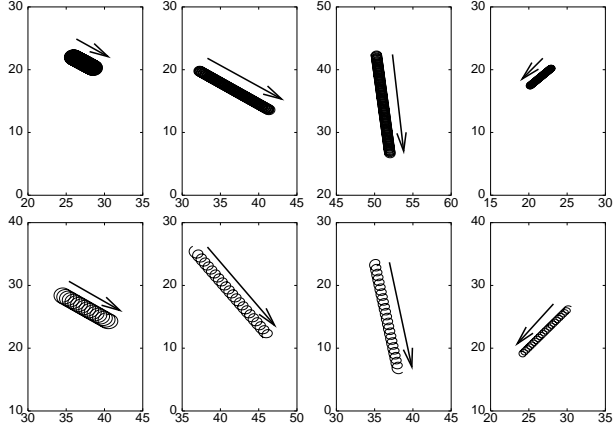}
\end{tabular}
\caption{Tip trajectories in the presence of an electric field 
for $a$ equal (left to right) to 0.85, 0.91, 0.93 and 0.99 for E=0.03 (top) 
and E=0.003 bottom (Other parameter values are $b=0.13$ and $\epsilon=0.02$).
 The trajectories were recorded for 100. time units (tu) (top) and 400 tu (bottom) and are represented in boxes of same spatial extension. The arrows show the drift direction. For parameter values $a=0.85$ and $a=0.99$, the drift velocity behaves almost linearly with the strength of the field (this was checked by performing two simulations, see also fig. \ref{driftcoef} (one should note that this is enough to check wether the linear regime is reached since an additional point is $v=0$, $E=0$) while in the cases $a=0.91$ and $a=0.93$ the behavior is still strongly non-linear.\label{traj_tip}}
\end{center}
\end{figure}

 I begin with describing the results of a numerical study of the spiral dynamics in the presence of an electric field using the Barkley model of an excitable medium (contrary to BZ systems, $u$ is the only diffusive variable): 
\begin{eqnarray}
\partial_t u&=& \frac{1}{\epsilon}f(u,v)+\Delta u+E\partial_x u,\label{reacdifu}\\
\partial_t v&=& g(u,v)\label{reacdifv},
\end{eqnarray}  
 where $f(u,v)=u(1-u)(u-\frac{v+b}{a})$,  $g(u,v)=u-v$ and $E$ is 
the electric field, directed along the $x$ axis. An homogeneous system modeled by these equations presents a single stable equilibrium point 
$u=v=0$ and a small perturbation of this equilibrium point can lead to a large excursion in phase space before return to equilibrium.  The diffusive coupling 
in Eq. \ref{reacdifu} allows the propagation of solitary waves in 1D. In 2D, in the absence of an electric field ($E=0$), a rich 
variety of wave propagation regime is observed depending on the values of the 
parameters\cite{Bar94}. I will focus here on  steadily rotating spirals 
and present results obtained along a line of equation $b=0.13$ in the parameter space of \cite{Bar94} (along this line no meandering spiral is observed). 

 In the presence of the electric field ($E\ne 0 $), the spiral tip drifts with a constant velocity (see figure \ref{traj_tip}) that depends on 
the parameter values and on the electric field strength. 
The drift velocity has  components perpendicular ($v_\perp$)
 and parallel ($v_\parallel$) to the field.  These velocities  vary linearly with the field in the small $E$ limit. This  allows to define the drift coefficient $\alpha_\perp=v_\perp/E$ and  $\alpha_\parallel=v_\parallel/E$\footnote{To avoid ambiguity, the direction of  $v_{\perp}$  is  defined  as the drift velocity in the direction of  $\mathbf{E}\times\mathbf{\omega}$ where $\mathbf{\omega}$ is the frequency of the spiral and $\mathbf{E}=(E,0)$.} in a weak field. 
 
\begin{figure}
\begin{center}
\includegraphics[width=0.5\textwidth]{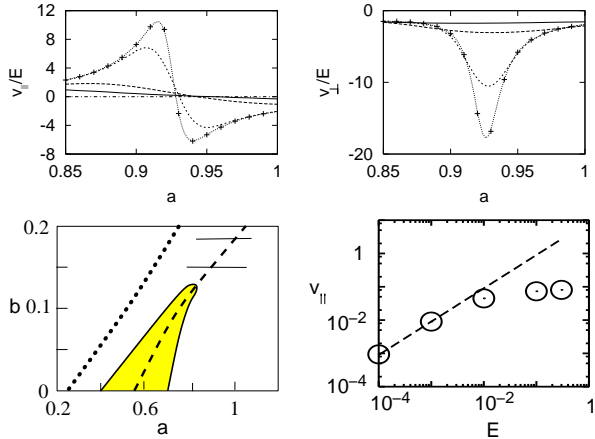}
\caption{\label{driftcoef} Parameter values are $\epsilon=0.02$ and $b=0.13$.
Top left: $v_\parallel/E$ for the following  values of the applied external field: 
0.3( solid), 0.1 (long dashed), 0.01 (short dashed), 0.001 (dotted), 0.0001(+), the dash-dotted line is the line of equation $v_\parallel/E=0$. Top right:  $v_\perp/E$ for the same values of the external field, $E$. Note that for both larger and smaller values of $a$ (not shown here) $v_\parallel/E$ and  $v_\perp/E$ are independent of $E$ for the $E$ range explored here. Note that the curves for $E=0.0001$ and $E=0.001$ are indistinguishable, indicating that for those values, the drift speed is linear in $E$. Hence the dotted curve also represent $\alpha_\perp$ and $\alpha_\parallel$. One should note the very different shape of the two dotted curves. Similar results (with a not so strong drift enhancement) were observed for higher values of $b$ and stronger damping of meander modes ($b=0.18$). Bottom left: schematic of the phase diagram in Barkley's parameter space. In the region left from the dotted line no spiral waves are observed. On the right hand side of this line, rotating and meandering spiral waves (gray region) are observed. The dashed line corresponds roughly to $\omega_m=\omega_t$ and the horizontal solid lines correspond to the values of $b$ (0.13 and 0.18) for which systematic simulations were performed. Bottom right: $v_\parallel$ as a function of $\log_{10}E$ for $a=0.92$ and $b=0.13$. $E$ is varying from 0.0001 to 0.3 and the dashed line is of equation $v_\parallel=9.3 v$.}
\end{center}
\end{figure}

Numerical  results (see Fig. \ref{driftcoef}) show that close 
to the transition from anti-parallel to parallel drift, the value of $E$ 
for which the linear regime is reached is much smaller than for 
the small $a$ and large $a$ region. In addition, close to this transition,
the dependence of the drift coefficients $\alpha_\parallel+i\alpha_\perp$ 
in $a$ is strongly non monotonous. When increasing $a$ there 
is a strong enhancement of the parallel drift coefficient 
followed by a relatively sharp transition from parallel to anti-parallel drift and finally the drift coefficients decrease rapidly. The behavior of the perpendicular drift coefficient is characterized by a strong enhancement close to the value of $a$ where the transition occurs. These results differ strongly from the results presented in \cite{Kri96}, where the drift speeds were computed using the same parameter values and an electric field of amplitude $E=0.3$: a monotonic variation of the drift speeds with the parameter values which is also observed here when using  high values of $E$ for which (see Fig. \ref{driftcoef}, solid line) the drift amplitude is no longer linear with the field amplitude.

 I now present an ODE model of spiral wave drift in an electric field that explains the strong enhancement of drift coefficients by the resonance of damped meander modes and translation modes.

 The ODE model is a modification of models \cite{Bar94,San2001} that reproduces the main feature of the spiral wave dynamics and has been modified  in order to take into account the effects  of a small electric field\cite{Blas}:
\begin{eqnarray}
\dot{r}&=&e^{i\phi}(R_0\omega_t-z)+\beta E\label{drift1}\\
\dot{z}&=&[\mu-i\omega_m-(1+i\alpha)|z^2|]z+\gamma E e^{-i\phi}\label{drift2}\\
\dot{\phi}&=&\omega_t \label{drift3}
\end{eqnarray} 
 where $r$ is the position of the tip in the complex plane ($(x,y)$ in the real plane correspond to $(Re,Im)$ in the complex plane)  and $z$ is a complex variable describing the meander in the frame rotating with the spiral and 
$\mu+i \omega_m$ is the eigenvalue associated with the meander mode and $E$ is the amplitude of the field directed along the real axis in the imaginary plane.  
In the absence of an electric field ($E=0$), for $\mu<0$, the steady state 
is obtained for $z=0$ and corresponds to the stationary rotating spiral 
$r=R_0 \exp(i\omega_t t)$ at frequency $\omega_t$. 
For $\mu>0$, the $z=0$ solution of Eq. \ref{drift2} is no longer stable and the $z$ variable undergoes  a Hopf bifurcation  that leads to the  modulation of the radius of curvature of the tip  with frequency $-\omega_m$ 
characteristic of meandering spirals (one should note that for $\omega_t=\omega_m$, the meander instability ($\mu>0$)leads to the drift of the rotating spiral with constant velocity \cite{Bar94} for $E=0$). 

I now give some rationale for the terms introduced in eqs. (\ref{drift1},\ref{drift2},\ref{drift3}). In eq. \ref{drift1}, the effect of the field can be either independent of the relative orientation of the spiral and the field (the $\beta E$ term) or depend on it. Nonetheless, the later case leads to a precessing term that once integrated over one spiral rotation vanishes at dominant order. Hence, I have chosen to add only the former term and to disregard the later that would have no effect at leading order.
The  $\gamma  E\exp(-i\phi)$  term in eq. \ref{drift2} has to take into account that the  field is constant in the laboratory frame while  eq. \ref{drift2} is expressed in the frame rotating with the spiral, that is  the laboratory frame rotated of $\phi$. Therefore it is necessary to add the $\exp(-i\phi)$ factor which expresses the fact that the meander mode is affected by the field in a way  that depends on the relative orientations of the spiral and of the electrical field (in the frame rotating with the spiral the field is rotating with a pulsation which is $-\dot{\phi}$).  No extra 
term has been added to Eq. \ref{drift3} (also expressed 
in the frame rotating with the spiral) since at dominant order the addition of a a term proportional to $E\exp(-i\phi)$ would only lead to  another constant drift term in Eq. \ref{drift1} similar to the $\beta E$ term.

  I focus here on the 
case where steadily rotating spirals are stable, (i.e. $\mu<0$ and $\omega_m$ is the frequency of the meander mode that can be computed by linear stability analysis\cite{Bar91a} of steady spirals) and describe the effects of a small field in this situation. 
  At the leading order in $E$ and $z$, Eq. \ref{drift2} has 
for solution $z=A \exp (-i\omega_t t)$ with $A=\gamma E/(-\mu+i(-\omega_t+\omega_m))$. Using this expression 
in Eq. \ref{drift1} leads  to a spiral tip  drifting with constant velocity:
\begin{equation}
\frac{v_d}{E}=\beta+ \frac{\gamma (\mu -i(\omega_t-\omega_m))}{\mu^2+(\omega_t-\omega_m)^2} \label{driftspeed}
\end{equation}

 This expression describes qualitatively the   
behavior of the spiral drift coefficients in the presence of an electric field  
as a function of $a$ presented in Fig. \ref{driftcoef} if the two following conditions are met. First close to the point where the transition from anti-parallel to parallel drift  occurs the meander frequency, $\omega_m$ becomes equal
 to the spiral frequency $\omega_t$ and $\mu$ is small (\textit{weakly} damped meander regime).  Second, the real part of $\gamma$ is much smaller than its imaginary part in this region. Else, the expression of Eq. \ref{driftspeed} can not reproduce qualitatively the behavior observed in fig. \ref{driftcoef}.  One can check this statement looking at the value of the drift  coefficient in the case where $\gamma=i\gamma_i$ is imaginary. In this case, the expression of the drift speed is  
 \begin{equation}
  \frac{v_d}{E}= \beta+\frac{\gamma_i(\omega_t-\omega_m)}{\mu^2+(\omega_t-\omega_m)^2} +i\frac{\gamma_i\mu}{\mu^2+(\omega_t-\omega_m)^2}
 \end{equation}
 which  reproduces well the behavior presented in fig. \ref{driftcoef} when  $\mu$ is small.
 
  It is also interesting to note that the  drift coefficient as expressed in eq. \ref{driftspeed} diverges if $\mu=0$ and $\omega_m=\omega_t$, that is at the codimension two point of Barkley phase space\cite{Bar94} (noted P here).  This seems unphysical since 
in the absence of an electric field one can see a perfectly stable 
spiral. Nonetheless, as shown in \cite{Bar94}, for  $\omega_m=\omega_t$ 
and $\mu>0$ the meandering spiral tip trajectory  is the one of a steadily drifting spiral with a constant finite speed even if $E=0$ (this situation corresponds to the case of an infinite drift coefficient). Hence it is not surprising that the drift coefficient should diverge when approaching P in the stable spiral region.

The following part of this paper will show that these requirements are met. In addition, the link between the drift of a spiral wave  and three dimensional instabilities of scroll waves will be discussed.

\begin{figure}
\begin{center}
\includegraphics[width=0.5\textwidth]{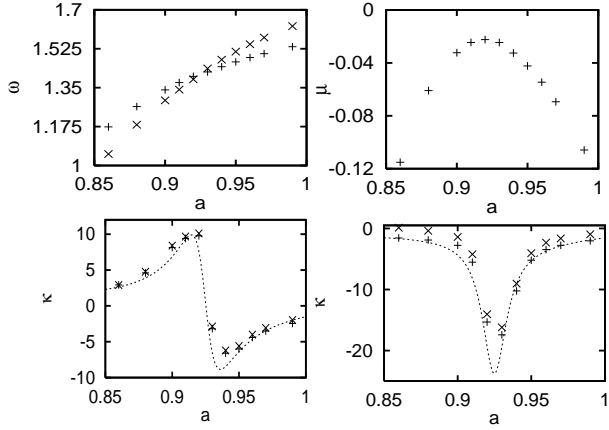}
\caption{\label{Fig_stablin}Top left: computed frequencies of the meander mode ($\times$) and of the translation mode ($+$) as a function of the control parameter $a$ ($\epsilon=0.02$ and $b=0.13$). $\omega_m$ and $\omega_t$ are equal for $a\approx 0.925$. Top right: growth rate of the meander mode as a function of $a$ (other parameters unchanged). The  maximal value of $\mu$ is reached for $a\approx 0.920$. Bottom left: curvatures of the real part of the translation branches  ($+$) and opposite of the curvature of the real part of the meander branches ($\times$) at $k_z=0$.  Bottom right: Computed curvatures of the imaginary part of the translation branches and opposite of the curvature of the imaginary part of the meander branches at $k_z=0$. For $a=0.92 \mbox{ and } 0.93$ the computed values of the curvatures present an uncertainty of order 1 due to the hybridization of meander and translation modes. Dashed line in bottom figures:  fit obtained using Eq. \ref{courbtrans} for the translation modes and a second order polynomial fit of $\gamma(a)$ and $\beta(a)$. Fit for the meander modes not shown here.}
\end{center}
\end{figure}

One can determine the values of $\omega_t$ and $\omega_m$ using the linear stability analysis of steady spiral previously described in \cite{Bar91a}
and extended to  scroll waves  in \cite{hh2000}. As seen in Fig. \ref{Fig_stablin}, $\omega_t$ and  $\omega_m$  become equal 
for $a\approx 0.925$ (i.e. close to the transition from anti-parallel to parallel drift). In addition, for this value of $a$, the damping of meander mode is relatively weak. Finally, taking  $\gamma=0.4i$ and $\beta=-0.36-1.3i$, 
the behavior of the drift coefficients is  well approximated by the expression of Eq. \ref{driftspeed} in the vicinity of  the transition. 

 Since the effects of a small electric field on a spiral are analogous to the effects of the slight curvature of a scroll wave, it is of interest to consider the three dimensional extension of this  model. In the following part of the paper, I will describe the results obtained using this simple model. The results presented here confirm that the low $k_z$ curvature of the translation mode\cite{hh2000} is equal to the drift coefficients. They will also  show that contrarily to what was hypothesized in \cite{Ara98}, the drift coefficients are not equal to the opposite of the low $k_z$ curvature of meander modes. 

 First,  consider a slightly curved spiral filament. The evolution equations of $u$ and $v$ are  the three dimensional analog of Eqs.~\ref{reacdifu}-\ref{reacdifv} in the case $E=0$. They can be 
rewritten using 
the coordinates $s,\ X,\ ,Y$ where $s$ is the curvilinear abscissa along the spiral filament\cite{Kee88}, $X$ and $Y$ are  the cartesian  coordinates in the plane  perpendicular to the filament (the $X$ axis being oriented along the normal to the filament). Using this coordinate system and assuming a slight curvature of the filament, Eq. \ref{reacdifv} is unchanged and  Eq. \ref{reacdifu} reads:
\begin{equation} 
\partial_t u =\frac{1}{\epsilon}f(u,v)+\Delta_{XY} u-\frac{1}{\rho-X}\partial_X u,\label{reacdifu_fil}
\end{equation}
where $\rho$ is the radius of curvature of the filament and $\Delta_{XY}$ is the two dimensional laplacian operator in the $XY$ plane. The $-(1/(\rho-X))\partial_X u$ term, at leading order in  $X/\rho$ is equal 
to $ 1/\rho$.  Hence in the new frame, close to the filament, Eq.~\ref{reacdifu} is affected by the curvature of the filament in the same way it is affected by an electric field $E=-1/\rho$. 

 Therefore,  the three dimensional extension of the  Barkley-Sandstede model for slightly curved filaments is given by  eq.
(\ref{drift1} \ref{drift2} \ref{drift3}) where $E$ is replaced by $-r''$ and 
 where $r''$ denotes the second order derivative of $r$ along the axis of the filament. For slightly curved filaments $r''$ is equal to the opposite of 
its curvature. A linear stability analysis of the tridimensional ODE model around the steady scroll wave ($z=0$, $r_0(z,t)=\exp(i\omega_t t)$) in the frame rotating with the spiral i.e. using a perturbation of the form $
r=r_0(z,t)+r_1 e^{i\phi}e^{\sigma t+ik_zz} \mbox{ and }
z=z_1 e^{\sigma t+ik_zz}$
leads to the following eigenvalue problem:
\begin{equation}
\sigma\left(
\begin{array}{c}
r_1\\
z_1
\end{array}
\right)
=
\left(
\begin{array}{cc}
-i\omega_t+\beta k_z^2 &-1\\
+\gamma k_z ^2 &\mu-i\omega_m
\end{array}
\right)
\left(
\begin{array}{c}
r_1\\
z_1
\end{array}
\right)
\label{stablin3D}
\end{equation}
 whose eigenvalues are at leading order in $k_z$:
\begin{eqnarray}
\sigma_t&=&-i\omega_t+\left(\beta+\frac{\gamma}{\mu+i(\omega_t-\omega_m)}\right)k_z^2\label{courbtrans}\\
\sigma_m&=&\mu-i\omega_m-\left(\frac{\gamma}{\mu+i(\omega_t-\omega_m)}\right)k_z^2\label{courbmeandre}
\end{eqnarray}
 These equations show that the curvature of the translation branch is equal to 
the opposite of the curvature of the meander mode shifted of $\beta$. It is also equal to  the drift speed (see Eq.\ref{driftspeed}).
 Hence, contrary to what was proposed by Aranson \textit{et al.}\cite{Ara98}, the curvature of the meander mode is not equal to  the drift coefficients. There is a shift (that can be relatively small when the parameter values are close  o the line where $\omega_m=\omega_t$) as shown in Fig. \ref{Fig_stablin}. This might come from the fact that the parameter $\beta$ was omitted in \cite{Ara98}.

 In addition, since the numerical linear stability analysis of scroll waves 
allows to determine $\sigma_t$, $\sigma_m$, $\mu$, $\omega_t$, and $\omega_m$
 for small values of $k_z$, one can extract from Eqs.\ref{courbtrans},\ref{courbmeandre} the expressions of the coefficients 
$\beta$ and $\gamma$:
\begin{eqnarray}
\beta&=&-(\kappa_m+\kappa_t)\label{betaa}\\
\gamma&=&\kappa_m (\mu+i(\omega_t-\omega_m))\label{gammaa}\\
\kappa_m&=& -(u_m \bullet \tilde{u}_m)/((u_m \bullet \tilde{u}_m)+(v_m \bullet \tilde{v}_m))\\
\kappa_t&=& -(u_t \bullet \tilde{u}_t)/((u_t\bullet \tilde{u}_t)+(v_t \bullet \tilde{v}_t))
\end{eqnarray} 
 where $\kappa_m$ and $\kappa_t$ are the respective curvatures of meander 
and translation branches at $k_z=0$ and can be expressed as functions of 
$(u_m,v_m)$, $(u_t,v_t)$ the meander and translation  modes,  $(\tilde{u}_m,\tilde{v}_m)$, $(\tilde{u}_t,\tilde{v}_t)$ are the corresponding adjoint modes that can be computed\cite{hh2002} and $(\bullet)$ denotes the usual scalar product in the frame rotating with the spiral. 

This method in the vicinity
of $\omega_m=\omega_t$  results in great variations in the values of 
$\gamma$ since it is very sensitive to possible inaccuracies. However, 
for values of $a$ away from the transition, one can compute with a 
good accuracy both $\beta(a)$ and $\gamma(a)$. They appear then to be 
smooth functions of $a$ that can be easily fitted by a second order in $a$ 
polynomial. For $a$ close to the transition  from parallel to anti-parallel drift, the polynomial fits of  $\beta(a)$ 
and $\gamma(a)$ takes values close to the ones used previously to fit 
the drift speed and using  the polynomial fit in Eqs. \ref{courbtrans}-\ref{courbmeandre} one  reproduces with a good accuracy the results of 
the long wavelength linear stability analysis of scroll 
waves (see Fig. \ref{Fig_stablin}).  One should also note that Eq. \ref{gammaa} together with the results of the numerical three dimensional linear stability analysis od scroll waves
give some rationale for $\gamma$ being imaginary close to the transition 
 from anti-parallel to parallel drift. Indeed, close to the transition $\kappa_m$ is mainly directed along the imaginary axis in the complex plane and $\omega_m\approx \omega_t$, those two results lead to $\gamma$ being directed almost 
along the imaginary axis.

 To conclude,
 using smoothly varying coefficients $\beta$ and $\gamma$, 
this model quantitatively
reproduces the results obtained numerically when considering both the drift of a spiral wave in the presence of an external field and the long
wavelength instabilities of scroll waves (meander and curvature). The reduced 
model brings  some clarification on the mechanism of the 
drift of a spiral wave  in the presence of an electric field, showing that the change in the sign of 
drift velocity parallel to the electric field can be attributed to a 
resonance between  meander and translation modes and that this resonance 
leads to a strong  increase in the drift coefficients. One should also note that despite the results presented here are formally  valid for $\mu<<1$, the resonance described here influences the drift speed as a function of parameters for a wider parameter regime and its effects can be observed rather far away from the Barkley's codimension 2 point (see fig. 2 a (inset)) as long as meander modes are not too  damped. I also hope those results can be checked experimentally in BZ system where meandering of spiral waves has been well characterized\cite{Ouyang}.

\acknowledgments
 I wish to thank Vincent Hakim and Blas Echebarria for very fruitful discussions. I am also grateful to Wouter-Jan Rappel and Vincent Hakim for useful comments on early versions of this manuscript. 
This work was supported in part by the NFS sponsored Center for Theoretical  Biological Physics (grant 0225630).

\end{document}